\documentclass[12pt,a4paper,titlepage]{article}

\usepackage{graphicx} 
\usepackage{amsmath,amssymb}
\usepackage{slashed}
\usepackage{xcolor}
\usepackage{hyperref}

\newcommand{\der}{\mathrm{d}}

\newcommand{\veck}{{\bf k}}
\newcommand{\veckj}{{\bf k}_{J}}
\newcommand{\veckji}{{\bf k}_{J,i}}
\newcommand{\veckjone}{{\bf k}_{J,1}}
\newcommand{\veckjtwo}{{\bf k}_{J,2}}

\newcommand{\non}{\nonumber\\}

\newcommand{\avgcos}{\langle \cos \varphi \rangle}
\newcommand{\avgcostwo}{\langle \cos 2 \varphi \rangle}
\newcommand{\avgcosn}{\langle \cos n \varphi \rangle}

\newcommand{\kina}{$\sqrt{s}=7$ TeV, $|\veckjone|=|\veckjtwo|=35$ GeV}
\newcommand{\kinb}{$\sqrt{s}=14$ TeV, $|\veckjone|=|\veckjtwo|=35$ GeV}
\newcommand{\kinc}{$\sqrt{s}=14$ TeV, $|\veckjone|=|\veckjtwo|=20$ GeV}
\newcommand{\kind}{$\sqrt{s}=14$ TeV, $|\veckjone|=|\veckjtwo|=10$ GeV}

\newcommand{\fig}{Fig.}
\newcommand{\eq}{Eq.}

\begin{document}
	\begin{titlepage}
		
	\begin{flushright}
		\begin{tabular}{l}
			LPT-ORSAY-15-52 
		\end{tabular}
	\end{flushright}
	\vspace{1.5cm}

	\begin{center}
		{\LARGE \bf Evaluating the double parton scattering contribution to Mueller-Navelet jets production at the LHC}
		\vspace{1cm}
		
		\renewcommand{\thefootnote}{\alph{footnote}}
		
		{\sc B.~Duclou\'e}$^{1,2}$,
		{\sc L.~Szymanowski}$^{3}$,
		{\sc S.~Wallon}$^{4,5}$.
		\\[0.5cm]
		\vspace*{0.1cm} ${}^1${\it Department of Physics, University of Jyv\"askyl\"a,		P.O. Box 35, 40014 University of Jyv\"askyl\"a, Finland} \\[0.2cm]
		\vspace*{0.1cm} ${}^2${\it Helsinki Institute of Physics, P.O. Box 64, 00014 University of Helsinki,	Finland} \\[0.2cm]
		\vspace*{0.1cm} ${}^3${\it National Centre for Nuclear Research (NCBJ), Warsaw, Poland} \\[0.2cm]
		\vspace*{0.1cm} ${}^4${\it LPT, Universit{\'e} Paris-Sud, CNRS, 91405, Orsay, France} \\[0.2cm]
		\vspace*{0.1cm} ${}^5${\it UPMC Univ. Paris 06, facult\'e de physique, 4 place Jussieu, 75252 Paris Cedex 05, France} \\[1.0cm]
		
		\vskip2cm
		{\bf Abstract:\\[10pt]} \parbox[t]{\textwidth}{
		We propose a model to study the importance of double parton scattering (DPS) in Mueller-Navelet jets production at the LHC which is consistent with the BFKL framework used to compute the single parton scattering contribution to this process. We study this model in kinematics corresponding to existing and possible future measurements at the LHC and estimate the importance of this DPS contribution on relevant observables for this process, namely the cross section and the azimuthal correlation of the jets.}
		\vskip1cm
	\end{center}
		
		\vspace*{1cm}
	\end{titlepage}

	\section{Introduction}
	
	The study of the production of two jets separated by a large interval of rapidity at hadron colliders was suggested by Mueller and Navelet~\cite{Mueller:1986ey} as a possible test of QCD in the high energy limit, which should be described by the Balitsky-Fadin-Kuraev-Lipatov (BFKL) approach~\cite{Fadin:1975cb,Kuraev:1976ge,Kuraev:1977fs,Balitsky:1978ic}. This pioneering study showed that, when the longitudinal momentum fractions of both jets are kept fixed, a BFKL calculation at leading logarithmic (LL) accuracy predicts a strong rise of the cross section when the rapidity separation between the jets increases, which is not expected in a fixed order approach. Later, another observable was proposed: the azimuthal correlation of the jets~\cite{DelDuca:1993mn,Stirling:1994he}. The physical picture is that in a pure leading order collinear treatment the two jets would be emitted exactly back to back, while the multiple emissions of gluons in the rapidity interval between the jets in the BFKL framework should 
lead to a loss of angular memory and so a stronger decorrelation.
	
	Recently the azimuthal correlations of Mueller-Navelet jets were measured for the first time at the LHC by the CMS collaboration~\cite{CMS-PAS-FSQ-12-002}. It was found that this measurement can be described with reasonable accuracy~\cite{Ducloue:2013bva} by a next-to-leading logarithmic (NLL) BFKL calculation supplemented by the use of the Brodsky-Lepage-Mackenzie procedure~\cite{Brodsky:1982gc} adapted to BFKL dynamics~\cite{Brodsky:1996sg,Brodsky:1997sd,Brodsky:1998kn,Brodsky:2002ka} to fix the renormalization scale (see also~\cite{Angioni:2011wj,Hentschinski:2012kr,Hentschinski:2013id,Caporale:2015uva}). However, at high energies and low transverse momenta, which is the region where BFKL effects are expected to be enhanced, parton densities can become large enough that contributions where several partons from the same incoming hadron take part in the interaction could become important. These multipartonic interactions (MPI) would make the comparison of BFKL calculations with experimental data much more 
difficult. The goal of the 
present paper is to propose a model to study MPI effects in this process and to evaluate their importance in the kinematics of the CMS study and of potential future studies at the LHC. We will restrict ourselves to the case where there are at most two scattering subprocesses and where both these scatterings are hard. This is generally known in the literature as double parton scattering (DPS). We note that some work was already done in this direction, where it was found that the DPS contribution could become more important than the single parton scattering (SPS) one in some kinematics~\cite{Maciula:2014pla}. However the DPS contribution was evaluated in the collinear framework which makes the comparison with a BFKL calculation questionable. Here we propose a model to evaluate the DPS contribution in a way which is consistent with the BFKL framework.
	
	\section{Formalism}
		
	\subsection{BFKL calculation}
	
	A detailed description of the BFKL calculation used to compute the SPS contribution to Mueller-Navelet jets production can be found in Refs.~\cite{Colferai:2010wu,Ducloue:2013hia,Ducloue:2013bva,Ducloue:2014koa}. For similar treatment see ~\cite{Caporale:2012ih,Caporale:2013uva,Caporale:2014gpa,Celiberto:2015yba}
	
	Here we will only illustrate a few key points that will serve as a basis to build our DPS model.
		
\begin{figure}[h]
		\centering\includegraphics[height=8cm]{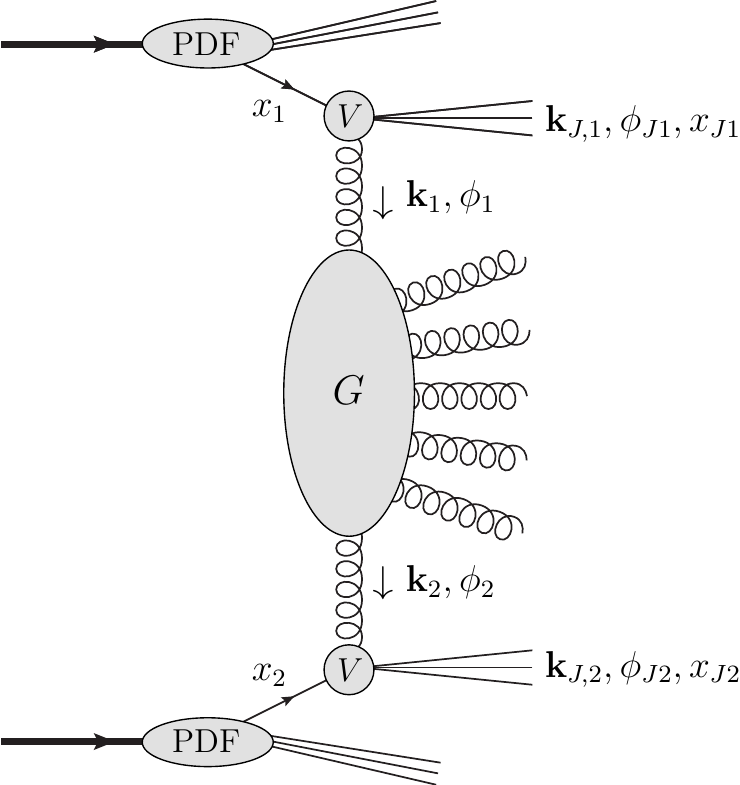}
		\caption{Kinematics of the SPS contribution in the BFKL approach.}
		\label{fig:SPS}
	\end{figure}

	A diagrammatic representation of this process is shown on \fig~\ref{fig:SPS}. Both incoming protons radiate a collinear parton which couples to the BFKL Green's function $G$, describing the multiple emission of untagged gluons, and the corresponding outgoing jet via the jet vertex. The differential cross section for jets with transverse momenta $\veckji$ and rapidities $y_{J,i}$ reads
	\begin{equation}
	\frac{\der \sigma}{\der |\veckjone| \der |\veckjtwo| \der y_{J,1} \der y_{J,2}} = \int \der \phi_{J,1} \der \phi_{J,2}\int\der^2\veck_1 \der^2\veck_2 \Phi(x_{J,1},-\veck_1)G(\veck_1,\veck_2,\hat s)\Phi(x_{J,2},\veck_2) \, ,
	\label{eq:sigma_BFKL}
	\end{equation}
	where the impact factors $\Phi$ are a convolution of the jet vertex $V$ with the parton distribution function (PDF) $f$:
	\begin{align}
	\Phi(x_{J,i},\veck_i) =& \int \der x_i f(x_i) V(\veck_i,x_i).
	\end{align}
	The leading order jet vertex is
	\begin{equation}
	V_{\rm a}^{(0)}(\veck_i,x_i)=\frac{\alpha_s}{\sqrt{2}}\frac{C_{A/F}}{\veck_i^2}\delta\left(1-\frac{x_{J,i}}{x_i}\right)|\veckji|\delta^{(2)}(\veck_i-\veckji) \, ,
	\label{eq:LO_vertex}
	\end{equation}
	where $C_A=N_c=3$ is to be used when the proton radiates a gluon and $C_F=(N_c^2-1)/(2N_c)=4/3$ when it radiates a quark.

	\subsection{DPS model}
	
\begin{figure}[h]
		\centering\includegraphics[height=5cm]{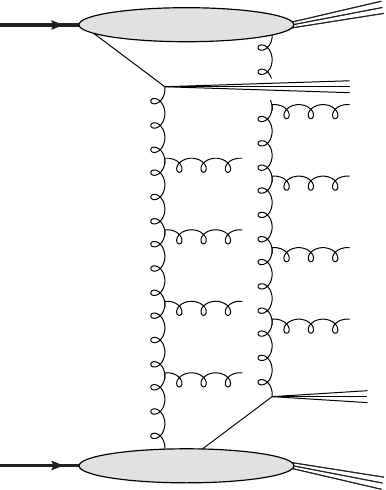}
		\caption{The DPS contribution.}
		\label{fig:DPS}
	\end{figure}

	To model the DPS contribution to this process, we want to stay as close as possible to the spirit of the BFKL approach described in the previous section to avoid discrepancies which could arise due to using different frameworks. We illustrate the DPS contribution on \fig~\ref{fig:DPS}. In this model, each jet is emitted by a different BFKL-like ladder. Each incoming proton emits two partons: one collinear, which enters the jet vertex ``close'' to its fragmentation region, and a gluon with nonzero transverse momentum which initiates the ladder attached to the other proton.
	
	In practice, to compute this DPS contribution, one would need to introduce some kind of ``hybrid'' double parton distributions, related to the probability to emit both a collinear parton and a gluon with some transverse momentum. Since at the moment almost nothing is known about such distributions, we use instead the simple factorized ansatz to compute the DPS contribution according to
	
	\begin{equation}
		\sigma_{\rm DPS}=\frac{\sigma_{\rm fwd} \sigma_{\rm bwd}}{\sigma_{\rm eff}} \, ,
		\label{eq:sigma_DPS}
	\end{equation}
	where $\sigma_{\rm fwd (bwd)}$ is the inclusive cross section for one jet in the forward (backward) direction and $\sigma_{\rm eff}$ is a phenomenological quantity related to the density of the proton in the transverse plane. According to measurements at the Tevatron~\cite{Abe:1993rv,Abe:1997xk,Abazov:2009gc,Abazov:2014fha} and at the LHC~\cite{Aad:2013bjm,Chatrchyan:2013xxa}, $\sigma_{\rm eff}$ should be of the order of 15 mb but there is some discrepancy between these measurements. To account for this uncertainty we will vary $\sigma_{\rm eff}$ between 10 and 20 mb in our calculation (note that this already amounts to an uncertainty of a factor 2 on the DPS cross section).
	\begin{figure}[h]
		\centering
		\includegraphics[height=5cm]{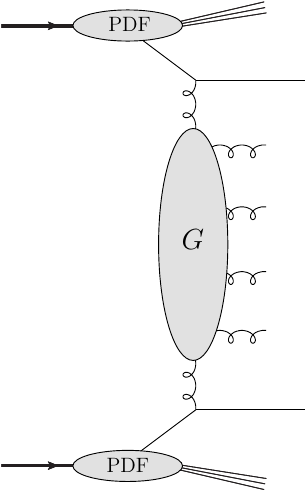}
		\hspace{3cm}
		\includegraphics[height=5cm]{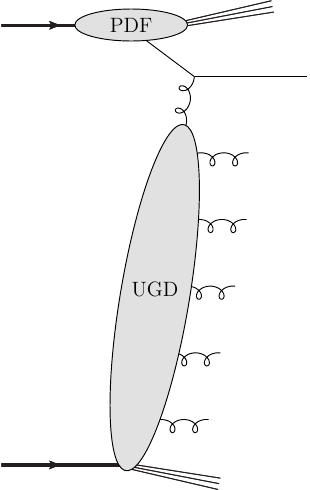}
		\caption{Left: Mueller-Navelet jets production at LL accuracy. Right: Inclusive forward jet production.}
		\label{fig:onejet}
	\end{figure}
	By using the approximation in \eq~(\ref{eq:sigma_DPS}), we have reduced the problem to the calculation of the cross section for the production of one jet in the forward (or backward) direction.
	
\begin{figure}[h]
		\centering\includegraphics[width=12cm]{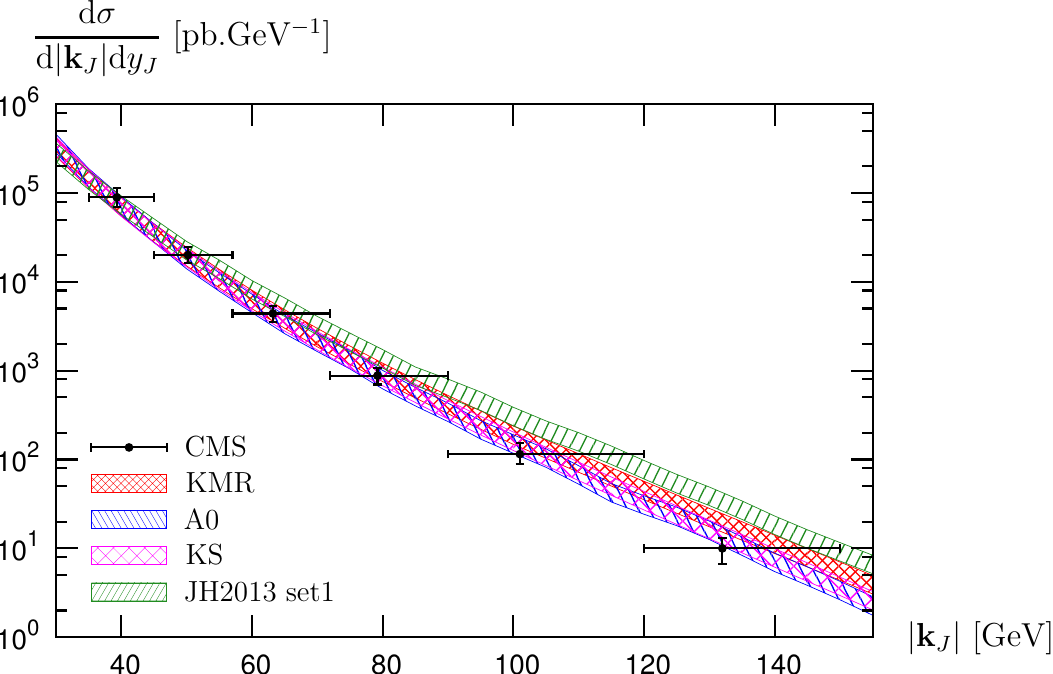}
		\caption{Comparison of CMS data at $\sqrt{s}=7$ TeV for inclusive forward jet cross section in the range $3.2<|y_J|<4.7$ with estimates based on several UGD parametrizations (see text for references) after fixing the parameter $K$.}
		\label{fig:sigma_onejet}
	\end{figure}

	To compute this cross section we start from the BFKL cross section for dijet production in \eq~(\ref{eq:sigma_BFKL}) in the LL approximation, which is illustrated on \fig~\ref{fig:onejet}~(L). Our goal is to compute the cross section for one forward jet, as shown on \fig~\ref{fig:onejet}~(R). For this we replace the part of the diagram containing the coupling to the lower proton, the lower jet vertex and the Green's function by an unintegrated gluon distribution (UGD), which couples directly to the lower proton and describes the multiple gluon emission between this proton and the jet. For simplicity we use the LO jet vertex as written in \eq~(\ref{eq:LO_vertex}), therefore the transverse momentum probed in the UGD is the one of the produced jet and its longitudinal momentum fraction value $x$ is given by the condition for the outgoing parton, which will then form the jet, to be on-shell. Unintegrated gluon distributions are much less known than usual collinear PDFs. Several models have been proposed, see 
for example Refs.~\cite{Kwiecinski:1997ee,GolecBiernat:1999qd,Kimber:2001sc,Hansson:2003xz,Kutak:2004ym,Jung:2004gs,Kutak:2012rf,Hautmann:2013tba}, but they can differ significantly in some phase space regions. In particular, we observe that for inclusive forward jet production the normalization depends strongly on the parametrization chosen. For this reason we write the differential cross section in $k_T-$factorization~\cite{Cheng:1970ef, FL, GFL, Catani:1990xk, Catani:1990eg, Collins:1991ty, Levin:1991ry}
as
	\begin{equation}
		\frac{\der \sigma}{\der |\veckj| \der y_J}=K \, \frac{\alpha_s}{|\veckj|} \, x_J \left( C_F \, f_q(x_J) + C_A \, f_g(x_J) \right) \mathcal{F}_g\left(\frac{\veckj^2}{s \, x_J},|\veckj|\right) \, ,
		\label{eq:sigma_onejet}
	\end{equation}
	where the unintegrated gluon distribution is normalized according to
	\begin{equation}
		\int^{Q^2} \der\veck^2 \mathcal{F}_g(x,|\veck|) = x f_g(x,Q^2) \, ,
	\end{equation}
	and $K$ is a dimensionless normalization factor which we will fix independently for each UGD parametrization with LHC data as explained in the following.
	The cross section for one forward jet production as a function of the transverse momentum of the jet was measured by the CMS collaboration at a center of mass energy of 7 TeV in the range $3.2<|y_J|<4.7$~\cite{Chatrchyan:2012gwa}.
	For each parametrization, we determine the range of $K$ compatible with the CMS measurement in the lowest transverse momentum bin. We choose to fit data only in the first bin because none of the parametrizations can describe perfectly the data over the whole $p_T$ range and in the following we will be interested in rather low transverse momenta. The comparison with CMS data after fixing $K$ according to this procedure is shown on \fig~\ref{fig:sigma_onejet} for several choices of UGD parametrizations: KMR~\cite{Kimber:2001sc}, A0~\cite{Jung:2004gs}, KS~\cite{Kutak:2012rf} and JH2013~\cite{Hautmann:2013tba}. We see that all models describe reasonably well the trend of the data.
	
	\section{Results}
	
	Now that we have all the necessary ingredients to compute both the SPS and the DPS contributions to this process, we are ready to estimate their relative importance (in the following we will neglect any possible interference between them). First we will compare the magnitude of the cross sections for both contributions. Then we will study how the DPS contribution could affect the azimuthal correlations of the jets. We will focus on four choices of kinematical cuts:
	\begin{itemize}
		\item \kina,
		\item \kinb,
		\item \kinc,
		\item \kind.
	\end{itemize}
	The first choice is similar to the cuts used by the CMS analysis of azimuthal correlations of Mueller-Navelet jets at the LHC~\cite{CMS-PAS-FSQ-12-002}. It is important to evaluate the DPS contribution for these kinematics to make sure that the good agreement found in Ref.~\cite{Ducloue:2013bva} is not just a coincidence. The other three choices correspond to the higher center of mass energy that the LHC is expected to reach soon. The choice $|\veckjone|=|\veckjtwo|=35$ GeV allows us to compare with the results at 7 TeV to evaluate the importance of a change of $\sqrt{s}$. The last two choices correspond to lower transverse momenta at which measurements could become possible in the future. They are particularly relevant since MPI are expected to become more and more important at lower transverse momenta. The rapidities of the jets are restricted according to $0<y_{J,1}<4.7$ and $-4.7<y_{J,2}<0$. We use the MSTW 2008 parametrization~\cite{Martin:2009iq} for collinear parton densities. To estimate the 
uncertainty associated with the choice of the UGD parametrization needed to compute the DPS cross section, we use the same four parametrizations as in \fig~\ref{fig:sigma_onejet} and take the extreme values as an error band. In the case of the NLL BFKL calculation, for which a jet can be made of more than one parton, we use the anti-$k_t$ jet algorithm~\cite{Cacciari:2008gp} with a size parameter $R=0.5$.

	\subsection{Cross section}
	
	\begin{figure}[t]
		\centering
		\includegraphics{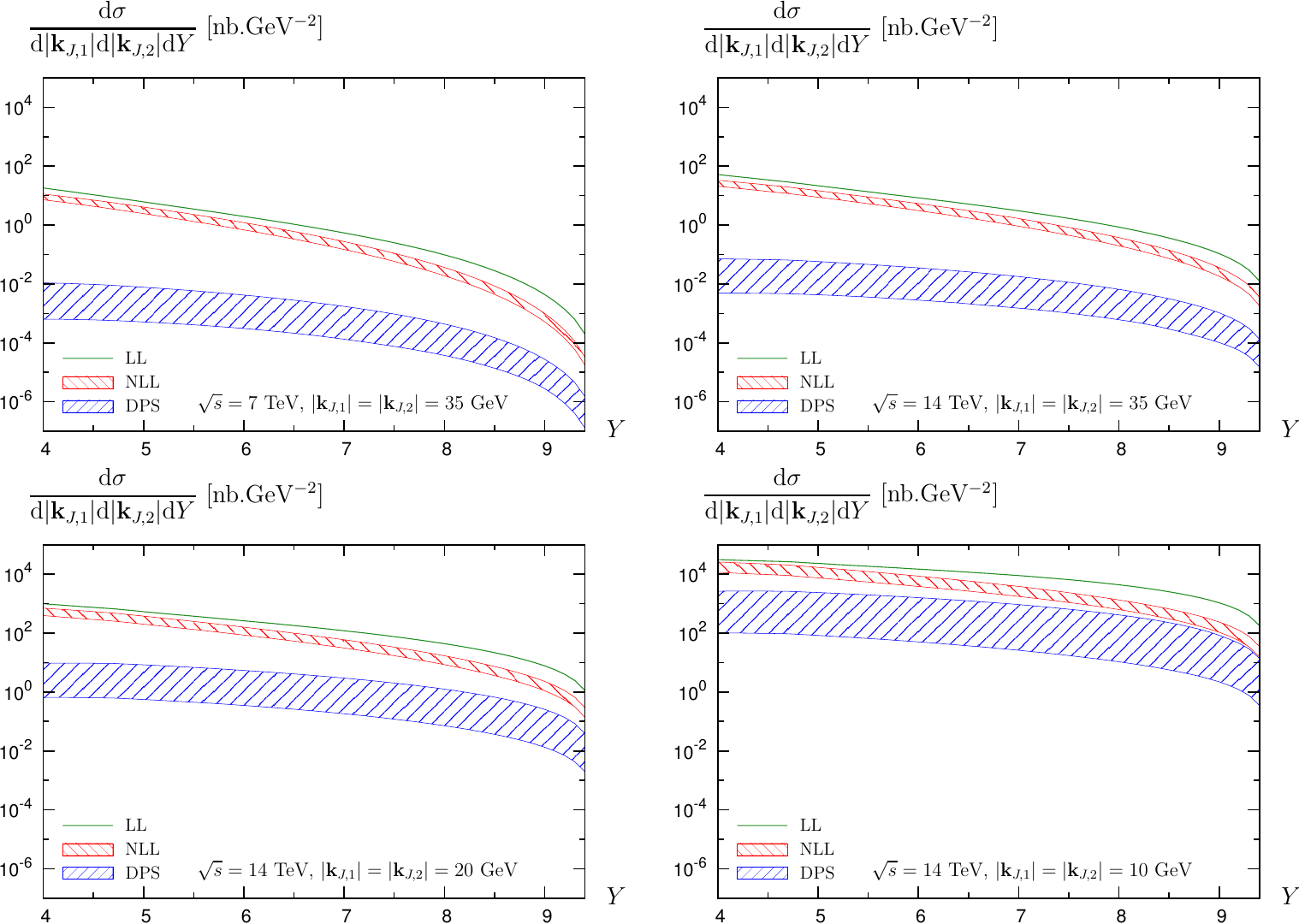}
		\caption{Comparison of the differential cross section obtained at LL (green) and NLL (red) accuracy in the BFKL approach and the DPS cross section (blue) for the four kinematical cuts described in the text.}
		\label{fig:sigma}
	\end{figure}
	
\begin{figure}[t]
		\centering
		\includegraphics{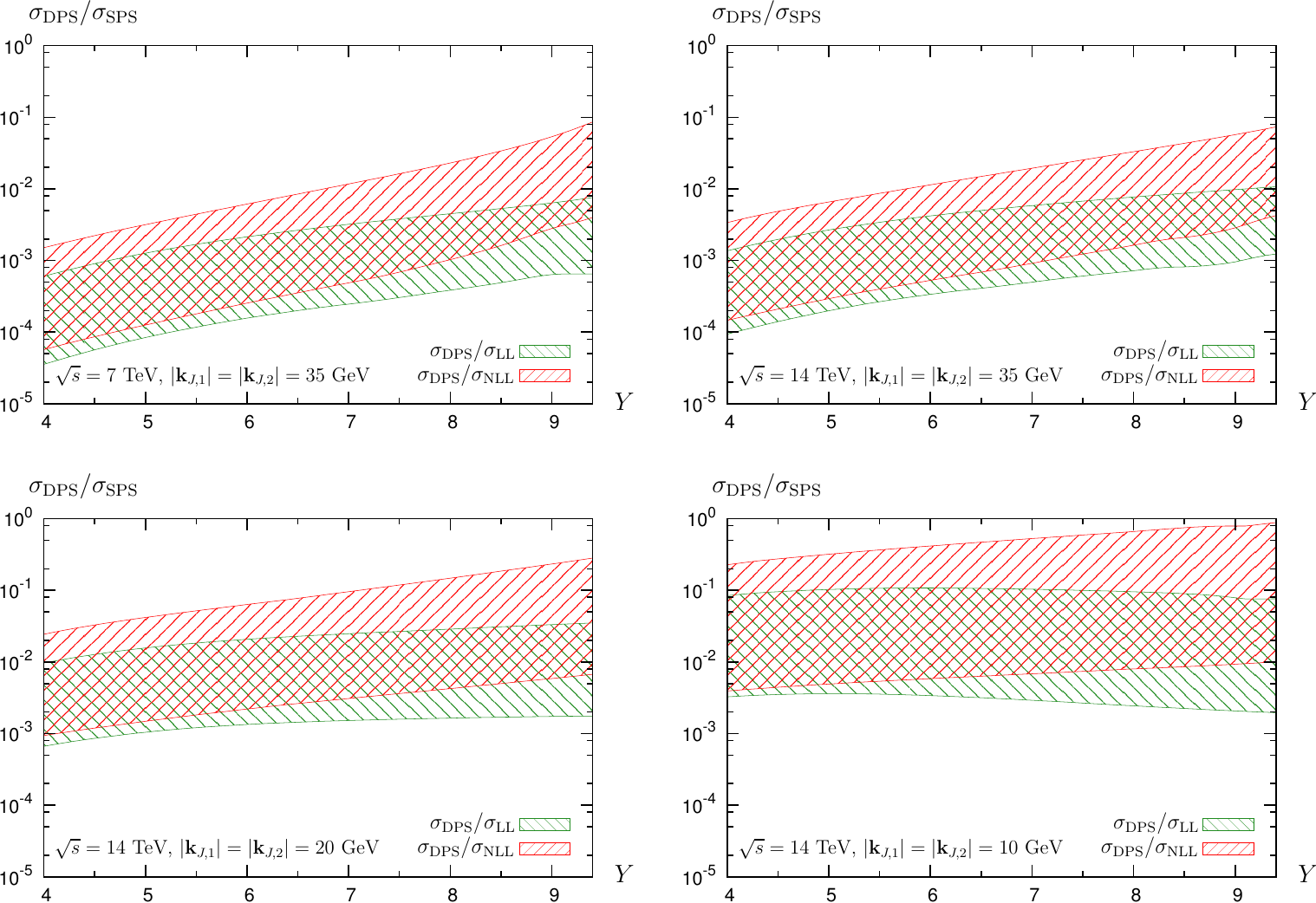}
		\caption{Ratio of the DPS cross section to the LL (green) and NLL (red) BFKL SPS cross section as a function of $Y$ for the four kinematical cuts described in the text.}
		\label{fig:ratio}
	\end{figure}

	In this section we compare the cross section obtained with the DPS model described in the previous section with the SPS contribution computed in the BFKL approach.
	On \fig~\ref{fig:sigma} we show, for the four kinematical cuts described above, the SPS cross section computed in the BFKL approach, either in a pure LL treatment or when including NLL corrections both to the Green's function~\cite{Fadin:1998py,Ciafaloni:1998gs} and to the jet vertex~\cite{Bartels:2001ge,Bartels:2002yj,Caporale:2011cc,Ivanov:2012ms,Colferai:2015zfa}, together with the DPS cross section.
	The first observation is that the uncertainty is much larger for the DPS cross section than for the SPS one. This uncertainty is due to the variation of $\sigma_{\rm eff}$ between 10 and 20 mb and the spread between the different UGD parametrizations. The cross section obtained in the three treatments increases at larger energy and smaller transverse momenta, especially for large rapidity separations. This is important because this means that a new analysis in the spirit of Ref.~\cite{CMS-PAS-FSQ-12-002} could provide much more statistics with the same luminosity. The DPS cross section is always smaller than both the LL and NLL BFKL calculations, but this difference reduces with increasing $Y$ and smaller transverse momenta. This can be seen more clearly on \fig~\ref{fig:ratio} where we show the ratio of the DPS cross section to the SPS one. This ratio grows faster with $Y$ when one treats the SPS contribution at NLL accuracy than at LL accuracy. This is due to the well-known fact that LL BFKL predicts a 
very strong rise of the partonic cross section with increasing rapidity separation between the jets. Note however that this LL calculation is not reliable anyway since for example it cannot describe CMS data on azimuthal correlations of the jets~\cite{CMS-PAS-FSQ-12-002} but we show it for illustrative purposes. It is important to note that for the first choice of kinematical cuts the DPS contribution is always smaller than the SPS one by at least one order of magnitude. As a consequence, the comparison made in Ref.~\cite{Ducloue:2013bva} with CMS data should be safe. The situation could be different for other kinematical conditions. For example, for \kind, at the edge of the uncertainty band the DPS contribution can become almost as large as the NLL BFKL cross section. However the uncertainty is again very large so this ratio is also compatible with much smaller values of the order of 1\%.

	\subsection{Azimuthal correlations}
	
	Even if the DPS cross section is always smaller than the SPS one in the kinematics under study here, there could be a measurable effect of DPS on more exclusive observables like the azimuthal correlation between the jets. Since our simple DPS model neglects any correlation between the jets, the inclusion of this contribution can only lead to a larger decorrelation than what is found in an SPS study, the size of this effect depending on the magnitude of the DPS cross section compared to the SPS one.
	
\begin{figure}[t]
		\centering
		\includegraphics{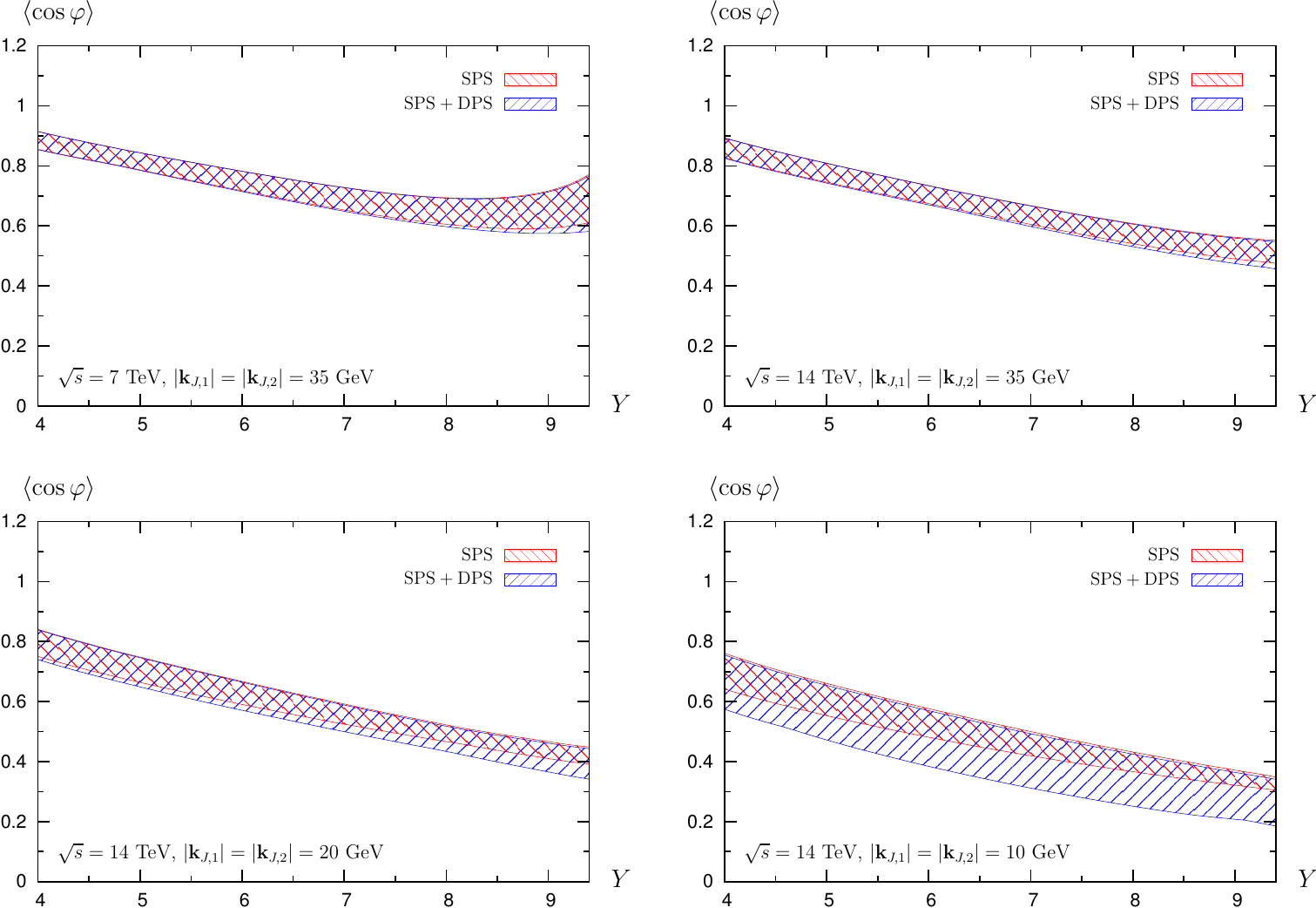}
		\caption{$\avgcos$ as a function of $Y$ without (red) and with (blue) the DPS contribution for the four kinematical cuts described in the text.}
		\label{fig:cos}
	\end{figure}

	To measure the azimuthal correlation between the jets, one can define the angular coefficients $\avgcosn$, with $\varphi=\phi_{J,1}-\phi_{J,2}-\pi$. In particular, $n=1$ corresponds to $\avgcos$ for which a value of 1 corresponds to always back-to-back jets while a value of 0 corresponds to completely uncorrelated jets (as we have for the DPS contribution). The correlation for all (SPS and DPS) events is given by
	\begin{align}
		\avgcosn&=\frac{\sigma_{\rm SPS}\avgcosn_{\rm SPS}+\sigma_{\rm DPS}\avgcosn_{\rm DPS}}{\sigma_{\rm SPS}+\sigma_{\rm DPS}} \non
		&=\frac{\avgcosn_{\rm SPS}}{1+\frac{\sigma_{\rm DPS}}{\sigma_{\rm SPS}}} \, ,
		\label{eq:cos_DPS}
	\end{align}
	where it is clear that
	\begin{equation}
		0<\avgcosn<\avgcosn_{\rm SPS} \, .
	\end{equation}
	On \fig~\ref{fig:cos} we show the first of these correlation coefficients, $\avgcos$, with and without the inclusion of DPS effects, for the four choices of kinematics described previously. In all cases the SPS and SPS+DPS calculations are compatible with each other. The reason is that, as shown on \fig~\ref{fig:ratio}, for some choices of the parameters the DPS cross section can be negligible compared to the SPS one. Even at the lower edge of the uncertainty band the deviation to SPS is not significantly larger than the uncertainty on the NLL BFKL calculation, except for the last choice of kinematics (\kind) at large rapidity separations.
	
	One can also study higher harmonics $\avgcosn$ with $n>1$. This allows us to compute ratios of the kind $\avgcostwo/\avgcos$ which have been found to be much less dependent on the choice of the scales than individual coefficients $\avgcosn$, both at LL and NLL accuracy~\cite{Vera:2006un,Vera:2007kn,Schwennsen:2007hs,Marquet:2007xx,Colferai:2010wu,Ducloue:2013hia}, and to show a good agreement with experimental data at NLL accuracy, even without using any scale-fixing procedure~\cite{Ducloue:2013bva}. These ratios are not affected by the uncorrelated DPS contribution under study here since, as can be seen from \eq~(\ref{eq:cos_DPS}), the quantity $1+\frac{\sigma_{\rm DPS}}{\sigma_{\rm SPS}}$ will disappear in any such ratio.
	
\begin{figure}[t]
		\centering
		\includegraphics{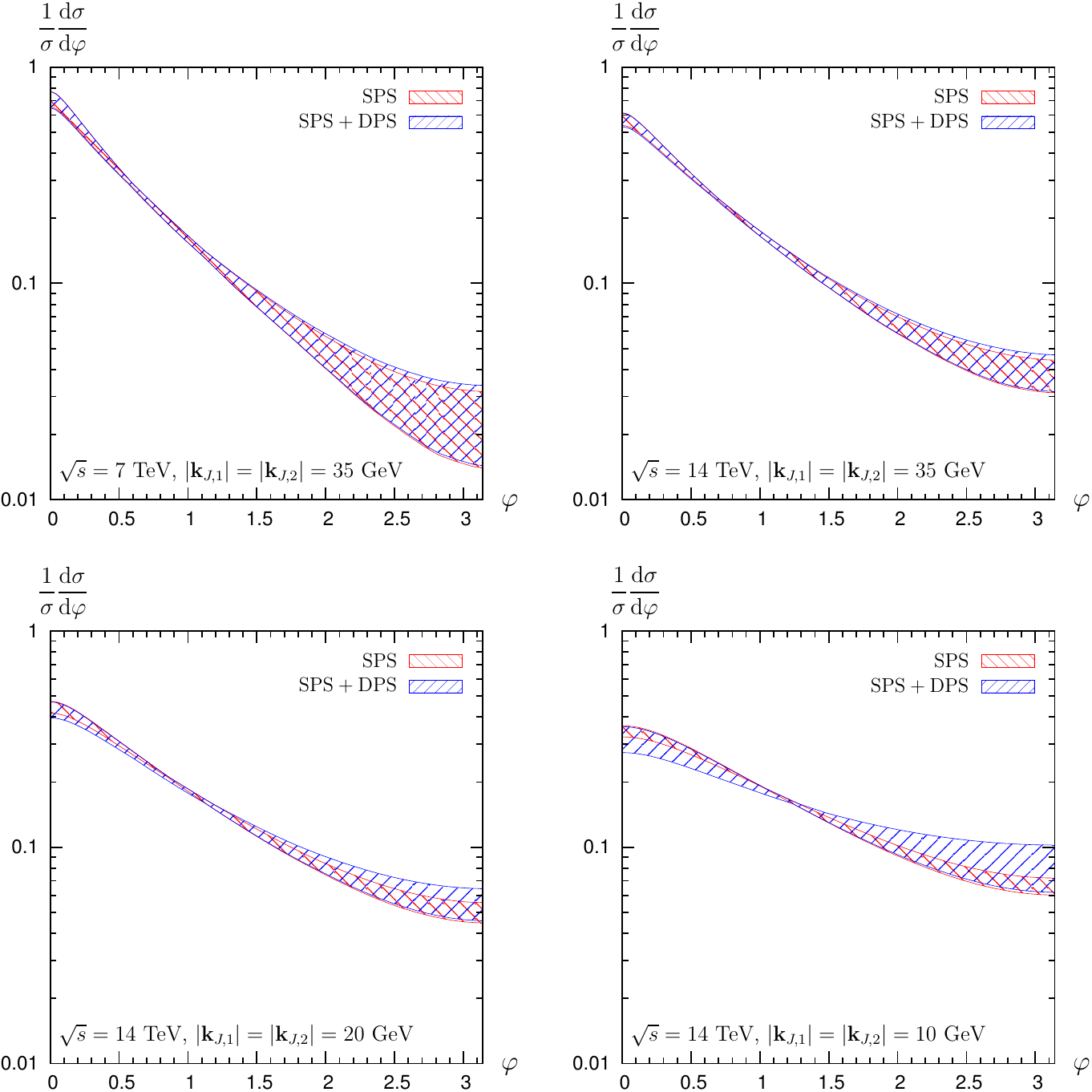}
		\caption{Azimuthal distribution $\frac{1}{\sigma}\frac{\der \sigma}{\der \varphi}$ integrated in the range $8<Y<9.4$ without (red) and with (blue) the DPS contribution, for the four kinematical cuts described in the text.}
		\label{fig:dist}
	\end{figure}

	Another observable related to the angular correlation of the jets is the azimuthal distribution $\frac{1}{\sigma}\frac{\der \sigma}{\der \varphi}$, which can be computed making use of the previously defined coefficients $\avgcosn$ according to
	\begin{equation}
		\frac{1}{{\sigma}}\frac{\der {\sigma}}{\der \varphi}=\frac{1}{2\pi}
		\left\{1+2 \sum_{n=1}^\infty \cos{\left(n \varphi\right)}
		\langle\cos{\left( n \varphi \right)}\rangle\right\} \, .
	\end{equation}
	Since in our model the DPS contribution produces uncorrelated jets, the corresponding azimuthal distribution is a constant. Similarly to the case of $\avgcos$ studied above, the total azimuthal distribution can be obtained by taking the average of the azimuthal distribution for the SPS and DPS contributions weighted by their respective cross section:
	\begin{align}
		\frac{1}{{\sigma}}\frac{\der {\sigma}}{\der \varphi}&=\frac{1}{\sigma_{\rm SPS}+\sigma_{\rm DPS}} \left(\frac{\der {\sigma_{\rm SPS}}}{\der \varphi}+\frac{\der {\sigma_{\rm DPS}}}{\der \varphi}\right) \non
		&=\frac{1}{\sigma_{\rm SPS}+\sigma_{\rm DPS}} \left(\frac{\der {\sigma_{\rm SPS}}}{\der \varphi}+\frac{\sigma_{\rm DPS}}{2\pi}\right) \, .
	\end{align}
	In \fig~\ref{fig:dist} we compare the azimuthal distribution obtained by taking into account both the SPS and DPS contributions with the results of the SPS calculation alone in the range $8<Y<9.4$. We observe that for the first three choices of kinematical cuts the maximal possible deviation from SPS induced by the DPS contribution is smaller than the uncertainty on the SPS calculation itself. On the contrary, for the choice \kind, the DPS contribution can lead to a significantly flatter distribution but there is an important overlap between the two treatments which are therefore still compatible with each other.
	
	\subsection{Determination of the DPS fraction from experimental data}
	
	In the previous sections we discussed the importance of double parton scattering for several observables in this process. However we found that our estimation is affected by a large uncertainty, which does not allow us to reach firm conclusions. To estimate the actual DPS contribution, one could try to extract the DPS fraction from future experimental data under the assumption that the DPS contribution is fully uncorrelated. As already explained above, such a DPS contribution would not affect conformal ratios like $\avgcostwo/\avgcos$. Therefore, if a future measurement shows that the BFKL calculation describes this observable well but overestimates significantly $\avgcos$, this could be interpreted as a large DPS fraction, which could be estimated by inverting the relation in \eq~(\ref{eq:cos_DPS}):
	\begin{equation}
		\frac{\sigma_{\rm DPS}}{\sigma_{\rm SPS}}=\frac{\avgcos_{\rm SPS}}{\avgcos}-1 \, ,
	\end{equation}
	where $\avgcos_{\rm SPS}$ is the value obtained with the BFKL calculation and $\avgcos$ is the value determined experimentally.

	\section{Conclusions}
	
	In this paper we proposed and studied a model to estimate the importance of double parton scattering in the production of two jets separated by a large interval of rapidity at hadron colliders. This simple model is based on the simple factorized ansatz which neglects any correlation between the jets emitted by two separate ladders. We estimated the uncertainty of our calculation by using different unintegrated gluon density parametrizations and by varying the value of $\sigma_{\rm eff}$ in a range compatible with experimental determinations of this quantity. It turns out that the resulting uncertainty on the DPS cross section is rather large. Still, this cross section is always smaller than the SPS one in the LHC kinematics we considered here. We also studied the impact of double parton scattering on the angular correlation between the jets. We found that the inclusion of this contribution leads to predictions still compatible with a NLL BFKL calculation including only single parton scattering. However, if 
one considers the set of parameters giving the largest DPS contribution, for low transverse momenta and large rapidity separations the effect of DPS can become larger than the uncertainty on the NLL BFKL calculation. Therefore in this region a more careful analysis or experimental data would be required to conclude. Finally, since our study relies on the relative size of the SPS and DPS cross sections, we would like to stress that a measurement of the absolute cross section for Mueller-Navelet jets production in kinematics similar to~\cite{CMS-PAS-FSQ-12-002}, where DPS effects are expected to be negligible, would be extremely valuable to confirm these results since it would provide a check of the SPS cross section normalization.
	
	\section*{Acknowledgements}
	
	We thank J.~Bartels, M.~Diehl, K.J.~Eskola, K.~Golec-Biernat, H.~Jung, T.~Kasemets, K.~Kutak, J.-P.~Lansberg, T.~Lappi, R.~Maciula, A.H.~Mueller and A. Szczurek for numerous discussions.
		
	B. D. is supported by the Academy of Finland, Project No.~273464.	
	The work of L. S. is supported by the Polish Grant NCN No. DEC-2011/01/B/ST2/03915, by the COPIN-IN2P3 Agreement and by the French grant ANR PARTONS No. ANR-12-MONU-0008-01.
	This work was done using computing resources from CSC -- IT Center for Science in Espoo, Finland.

\end{document}